\documentclass[12pt]{article}
\usepackage{epsfig}
\usepackage{graphicx}
\usepackage{amssymb}
\usepackage{amsmath}
\usepackage{amsthm}
\usepackage{amsfonts}
\usepackage{mathrsfs}
\usepackage[dvips]{color}
\usepackage{multirow}

\newcommand{\bM}{\mathbf{M}}

\newcommand{\cP}{\mathcal{P}}

\newcommand{\cT}{\mathcal{T}}

\newcommand{\be}{\begin{equation}}
\newcommand{\ee}{\end{equation}}
\newcommand{\bea}{\begin{eqnarray}}
\newcommand{\eea}{\end{eqnarray}}
\newcommand{\nn}{\nonumber}

\newcommand{\ed}{\end{document}}

\newcommand{\bi}{\begin{itemize}}
\newcommand{\ei}{\end{itemize}}

\newcommand{\bce}{\begin{center}}
\newcommand{\ece}{\end{center}}

\begin{document}

\title{Generalized Unitarity and Reciprocity Relations for $\cP\cT$-symmetric Scattering Potentials}

\author{Ali~Mostafazadeh\thanks{E-mail address:
amostafazadeh@ku.edu.tr, Phone: +90 212 338 1462, Fax: +90 212 338
1559}
\\
Department of Mathematics, Ko\c{c} University,\\
34450 Sar{\i}yer, Istanbul, Turkey}
\date{ }
\maketitle

\begin{abstract}
We derive certain identities satisfied by the left/right-reflection and transmission amplitudes, $R^{l/r}(k)$ and $T(k)$, of general $\cP\cT$-symmetric scattering potentials. We use these identities to give a general proof of the relations, $|T(-k)|=|T(k)|$ and $|R^r(-k)|=|R^l(k)|$, conjectured in [Z.~Ahmed, J.~Phys.~A \textbf{45} (2012) 032004], establish the generalized unitarity relation: $R^{l/r}(k)R^{l/r}(-k)+|T(k)|^2=1$, and show that it is a common property of both real and complex $\cP\cT$-symmetric potentials. The same holds for $T(-k)=T(k)^*$ and $|R^r(-k)|=|R^l(k)|$.
\end{abstract}

Recently there has been a growing interest in the scattering properties of complex potentials in general and $\cP\cT$-symmetric potentials in particular. This is especially motivated by the fact that these potentials are capable of supporting spectral singularities \cite{prl-2009,ss,pra-2011a,jpa-2012,ahmed-2012} and unidirectional reflectionlessness and invisibility \cite{invisible,pra-2013a}. Spectral singularities \cite{prl-2009} correspond to real poles of the left/right-reflection and transmission amplitudes, $R^{l/r}(k)$ and $T(k)$. They are of particular interest in optics, because they correspond to lasing at the threshold gain \cite{pra-2011a} while their time-reversal yields coherent perfect absorption (CPA) of the electromagnetic radiation \cite{longhi-2010,longhi-2010b}. Unidirectional reflectionlessness is characterized by the condition that $R^{l/r}(k)=0\neq R^{r/l}(k)$, and unidirectional invisibility is described through this condition together with the requirement of perfect transparency: $T(k)=1$. These phenomena are also of great interest, because they provide an interesting root towards designing one-way optical devices \cite{feng-yin}.

Real scattering potentials are incapable of supporting spectral singularities, CPA, and unidirectional reflectionlessness and invisibility, because they are constrained by the reciprocity and unitarity relations:
    \begin{align}
    &|R^l(k)|=|R^r(k)|,
    \label{e1}\\
    &|R^{l/r}(k)|^2+|T(k)|^2=1.
    \label{e2}
    \end{align}
While the first of these equations forbids unidirectional reflectionlessness (and hence invisibility), the second identifies the reflection and transmission coefficients, $|R^{l/r}(k)|^2$ and $|T(k)|^2$, with a pair of bounded functions that cannot possess singularities for real wavenumbers $k$.

Complex $\cP\cT$-symmetric scattering potentials $v(x)$, that by definition satisfy $v(-x)^*=v(x)$, can violate one or both of Eqs.~(\ref{e1}) and (\ref{e2}). Therefore, they provide a large class of models capable of displaying spectral singularities, CPA, and unidirectional reflectionlessness and invisibility. The fact that the spectral singularities of $\cP\cT$-symmetric potentials accompany their time-reversal dual \cite{jpa-2012} and that they possess the same symmetry as the equations governing unidirectional reflectionlessness and invisibility \cite{pra-2013a} have made them into a topic of intensive research.

In Ref.~\cite{Ge-2012} the authors derive the following $\cP\cT$-symmetric analog of the unitarity condition (\ref{e2}).
    \be
    |T(k)|^2\pm |R^l(k)R^r(k)|=1,
    \label{e3}
    \ee
where $\pm$ is to be identified with the sign of $1-|T(k)|^2$. This relation is most conveniently established using the properties of the transfer matrix \cite{jpa-2012}. This is the unique $2\times 2$ complex matrix $\bM(k)$ that connects the coefficients of the asymptotic solutions of the wave equation,
    \be
    \psi(x)\to A_\pm(k) e^{ikx}+B_\pm(k) e^{-ikx}~~~~{\rm for}~~~~x\to\pm\infty,
    \label{e4}
    \ee
according to
    \be
    \left[ \begin{array}{c} A_+(k)\\B_+(k)\end{array}\right]=\bM(k) \left[\begin{array}{c} A_-(k)\\B_-(k)\end{array}\right].
    \label{e5}
    \ee
The entries of $\bM(k)$ are related to the reflection and transmission amplitudes via
\cite{prl-2009}:
    \begin{align}
	&M_{11}(k)=T(k)-\frac{R^l(k) R^r(k)}{T(k)}, &&
	M_{12}(k)=\frac{R^r(k)}{T(k)}, && M_{21(k)}=-\frac{R^l(k)}{T(k)}, && M_{22}(k)=\frac{1}{T(k)}.
	\label{M-RT}
	\end{align}

In Ref.~\cite{ahmed-2012}, the author conjectures that the following identities hold for every $\cP\cT$-symmetric potential
    \begin{align}
    &|R^l(-k)|=|R^r(k)|,
    \label{e11}\\
    &|T(-k)|=|T(k)|,
    \label{e12}
    \end{align}
and uses a complexified Scarf II potential to provide evidence for this claim. The present investigation is motivated by the search for a general proof of these relations.

We begin our analysis by recalling that under the action of the space-reflection (parity) $\cP$ and time-reversal transformation $\cT$, the solutions of the wave equation $\psi(x)$ transform according to
    \begin{align}
    &\psi(x)\stackrel{\cP}{\longrightarrow}\psi(-x), &&
    &\psi(x)\stackrel{\cT}{\longrightarrow}\psi(x)^*.
    \label{P-T}
    \end{align}
It is not difficult to show that these equations together with (\ref{e4}) and (\ref{e5}) imply the following transformation rules for the transfer matrix \cite{longhi-2010b,jpa-2012}.
    \bea
    \bM(k)&\stackrel{\cP}{\longrightarrow}&\mathbf{\sigma}_1\bM(k)^{-1}\mathbf{\sigma}_1,
    \label{M-P}\\
    \bM(k)&\stackrel{\cT}{\longrightarrow}&\mathbf{\sigma}_1\bM(k)^{*}\mathbf{\sigma}_1,
    \label{M-T}\\
    \bM(k)&\stackrel{\cP\cT}{\longrightarrow}&\bM(k)^{-1*},
    \label{M-PT}
    \eea
where $k$ is taken to be real and $\mathbf{\sigma}_1$ is the first Pauli matrix (the $2\times 2$ matrix with vanishing diagonal entries and unit off-diagonal entries.) In terms of the entries of $\bM(k)$, these respectively take the form
    \begin{align}
    &M_{11}(k)\stackrel{\cP}{\longrightarrow}M_{11}(k),
    &&M_{12}(k)\stackrel{\cP}{\longleftrightarrow}-M_{21}(k),
    &&M_{22}(k)\stackrel{\cP}{\longrightarrow}M_{22}(k),
    \label{M-P2}\\
    &M_{11}(k)\stackrel{\cT}{\longleftrightarrow}M_{22}(k)^*,
    &&M_{12}(k)\stackrel{\cT}{\longleftrightarrow}M_{21}(k)^*,
    \label{M-T2}\\
    &M_{11}(k)\stackrel{\cP\cT}{\longleftrightarrow}M_{22}(k)^*,
    &&M_{12}(k)\stackrel{\cP\cT}{\longrightarrow}-M_{12}(k)^*,
    &&M_{21}(k)\stackrel{\cP\cT}{\longrightarrow}-M_{21}(k)^*
    \label{M-PT2}.
    \end{align}
The following are consequences of these relations.  
    \begin{itemize}
    \item If $v(x)$ is an even (real or complex) potential, $\bM$ in $\cP$-invariant, and (\ref{M-P2}) implies $M_{12}(k)=-M_{21}(k)$. In view of (\ref{M-RT}), this is equivalent to $R^l(k)=R^r(k)$.

    \item If $v(x)$ is a real potential, $\bM$ is $\cT$-invariant, and we find
    $M_{11}(k)^*=M_{22}(k)$ and $M_{12}(k)^*=M_{21}(k)$. Substituting (\ref{M-RT})
    in these relations leads to (\ref{e2}) and
        \be
        R^l(k)^*=-\frac{R^r(k)T(k)^*}{T(k)}.
        \label{e21}
        \ee
    Let $\lambda$, $\rho$, and $\tau$ be respectively the phase angles of $R^l(k)$, $R^r(k)$, and $T(k)$,
    so that
        \begin{align}
        & R^l(k)=e^{i\lambda(k)}|R^l(k)|,
        && R^r(k)=e^{i\rho(k)} |R^r(k)|,
        && T(k)=e^{i\tau(k)}|T(k)|.
        \label{e20}
        \end{align}
    Then, we can write (\ref{e21}) as
        \be
        R^l(k)^*=-e^{-2i\tau(k)}R^r(k).
        \label{e21b}
        \ee
    Taking the absolute-value of both sides of this relation gives (\ref{e1}). Now, suppose that $v(x)$ is not reflectionless for the wavenumber $k$. Then using (\ref{e1}) and (\ref{e20}) in (\ref{e21b}), we find
        \be
        \lambda(k)+\rho(k)=2\,\tau(k)+(2m+1)\pi,
        \label{e22}
        \ee
    where $m$ is an integer.

    \item If $v(x)$ is a $\cP\cT$-symmetric potential, $M_{11}(k)^*=M_{22}(k)$ and both $M_{12}(k)$ and $M_{21}(k)$ are purely imaginary. In view of (\ref{M-RT}) and (\ref{e20}), these imply the existence of integers $m_1$ and $m_2$ such that \cite{Ge-2012}
        \begin{align}
        &\lambda(k)=\tau(k)+\pi(m_1+\frac{1}{2}),
        \label{e31}\\
        & \rho(k)=\tau(k)+\pi(m_2+\frac{1}{2}),
        \label{e32}\\[4pt]
        &|T(k)|^2+e^{i\pi(m_1+m_2)}|R^r(k)R^l(k)|=1.
        \label{e33}
        \end{align}
    Clearly (\ref{e33}) coincides with (\ref{e3}). Another interesting observation is that adding Eqs.~(\ref{e31}) and (\ref{e32}) side by side gives
        \be
        \lambda(k)+\rho(k)=2\,\tau(k)+(m_1+m_2+1)\pi.
        \label{e22b}
        \ee
    This coincides with (\ref{e22}) whenever $m_1+m_2$ is even. In this case (\ref{e33}) reduces to $|T(k)|^2+|R^r(k)R^l(k)|=1$, and similarly to the case of real potentials, $|R^{l/r}|$ and $|T|$ are bounded-above by $1$. If $m_1+m_2$ is odd, (\ref{e33}) becomes $|T(k)|^2-|R^r(k)R^l(k)|=1$, $|R^{l/r}|$ and $|T|$ need not be bounded, and the potential can support spectral singularities.

    It is also worth mentioning that in light of (\ref{e33}), a $\cP\cT$-symmetric potential is unidirectionally or bidirectionally reflectionless at a wavenumber $k$ if and only if the transmission coefficient $|T(k)|^2$ is unity, i.e., $T(k)=e^{i\tau(k)}$.
    \end{itemize}

Another obvious consequence of Eqs.~(\ref{e4}) and (\ref{e5}) is the identity: $\bM(-k)=\mathbf{\sigma}_1\bM(k)\mathbf{\sigma}_1$, which means
	\begin{align}
	& M_{11}(-k)=M_{22}(k), && M_{12}(-k)=M_{21}(k).
	\label{e41}
	\end{align}
Writing these equations in terms of the reflection and transmission amplitudes, we find
	\begin{align}
	&R^l(-k)=-\frac{R^r(k)}{D(k)}, && R^r(-k)=-\frac{R^l(k)}{D(k)}, &&
    	T(-k)=\frac{T(k)}{D(k)},
    	\label{RRT}
    	\end{align}
where $D(k):=T(k)^2-R^l(k)R^r(k)$. The following are some notable implications of these relations.
    \begin{itemize}
    \item If $v(x)$ is a real potential, Eqs.\ (\ref{e2}) and (\ref{e21b}) hold, and we can use them together with (\ref{RRT}) to establish
        \begin{align}
        & D(k)=e^{2i\tau(k)}=\frac{T(k)}{T(k)^*},
        \label{D=}\\
        & R^{l/r}(-k)=R^{l/r}(k)^*,
        \label{e51}\\
        &T(-k)=T(k)^*.
        \label{e52}
        \end{align}
    Using the first two of these equations, we can respectively express the reciprocity and unitarity relations (\ref{e1}) and ~(\ref{e2}) as
        \bea
        &&|R^{l}(-k)|=|R^{r}(k)|,
        \label{e1-neq}\\
        &&R^{l/r}(k)R^{l/r}(-k)+|T(k)|^2=1.
        \label{e2-new}
        \eea
    Clearly, (\ref{e1-neq}) is identical to (\ref{e11}) and (\ref{e52}) implies (\ref{e12}). Therefore, Eqs.~(\ref{e11}) and (\ref{e12}) that are conjectured to hold for $\cP\cT$-symmetric potentials in Ref.~\cite{ahmed-2012} are indeed satisfied by real scattering potentials. Let us also note that for real potentials  (\ref{e11}) is just an alternative expression for the reciprocity relation (\ref{e1}).

    \item If $v(x)$ is a $\cP\cT$-symmetric potential, Eqs.~(\ref{e31}) -- (\ref{e22b}) hold, and it is easy to show that the last two of these relations imply (\ref{D=}). Using this equation in (\ref{RRT}), we find
        \begin{align}
        &R^{l/r}(-k)=-e^{2i\tau(k)}R^{r/l}(k), && T(-k)=T(k)^*.
        \label{e61}
        \end{align}
    If we take the absolute-value of both sides of these equations, we arrive at 
    Eqs.~(\ref{e11}) and (\ref{e12}). This provides a proof of the conjecture that every $\cP\cT$-symmetric scattering potential comply with (\ref{e11}) and (\ref{e12}), \cite{ahmed-2012}. As we explained above, these relations also hold for the real potentials and (\ref{e11}) is equivalent to the reciprocity relation (\ref{e1}). We may, therefore, refer to it as the ``reciprocity relation'' also for the complex $\cP\cT$-symmetric potentials.

    Next, we use the first equation in (\ref{e61}) together with (\ref{e20}), (\ref{e33}), and (\ref{e22b}) to compute:
        \bea
        R^l(k)R^l(-k)+|T(k)|^2&=&-e^{-2i\tau(k)}R^l(k)R^r(k)+|T(k)|^2\nn\\
        &=&-e^{i[\lambda(k)+\rho(k)-2\tau(k)]}|R^l(k)R^r(k)|+|T(k)|^2\nn\\
        &=&e^{i(m_1+m_2)\pi}|R^l(k)R^r(k)|+|T(k)|^2=1
        \eea
    Following the same approach, we also find $R^r(k)R^r(-k)+|T(k)|^2=1$. This leads us to the remarkable conclusion that, similarly to (\ref{D=}) and (\ref{e52}), (\ref{e2-new}) is also a property of real scattering potentials that is shared by complex $\cP\cT$-symmetric scattering potentials. Because for real potentials it coincides with the unitarity condition (\ref{e2}), we may view it as a generalized unitarity or pseudo-unitarity relation \cite{jmp-2004}.

    \end{itemize}

In conclusion, searching for a proof of the relations (\ref{e11}) and (\ref{e12}) and using the transformation properties of the transfer matrix of one-dimensional scattering theory, we have obtained a number of interesting identities for the reflection and transmission amplitudes of general $\cP\cT$-symmetric scattering potentials. Eqs.~(\ref{e11}) and (\ref{e12}) follow as immediate consequences of these identities and hold true also for real potentials. Perhaps more interestingly, we have found a particular form of the unitarity relation for real potentials, namely (\ref{e2-new}), that is also satisfied by $\cP\cT$-symmetric complex potentials. These observations uncover  interesting similarities between real and complex $\cP\cT$-symmetric scattering potentials.

\vspace{6pt}
\noindent {\em Acknowledgments:}  This work has been supported by  the Scientific and Technological Research Council of Turkey (T\"UB\.{I}TAK) in the framework of the project no: 112T951, and by the Turkish Academy of Sciences (T\"UBA).


\ed